# Experimental evidence for the influence of charge on the adsorption capacity of carbon dioxide on charged fullerenes

Stefan Ralser[1], Alexander Kaiser[1], Michael Probst[1], Johannes Postler[1], Michael Renzler[1], Diethard K. Bohme[2*] and Paul Scheier[1*]

[1] Institut für Ionenphysik und Angewandte Physik, Leopold-Franzens-Universität Innsbruck, Technikerstr. 25, 6020 Innsbruck, Austria.

[2] Department of Chemistry, York University, Toronto, ON, Canada M3J 1

*corresponding authors: Paul.Scheier@uibk.ac.at, dkbohme@yorku.ca

**We show, with both experiment and theory, that adsorption of $CO_2$ is sensitive to charge on a capturing model carbonaceous surface. In the experiment we dope superfluid helium droplets with $C_{60}$ and $CO_2$ and expose them to ionising free electrons. Both positively and negatively charged $C_{60}(CO_2)_n^{+/-}$ cluster ion distributions are observed with a high-resolution mass spectrometer and these show remarkable and reproducible anomalies in intensities that are strongly dependent on the charge. The highest adsorption capacity is seen with $C_{60}^+$. Complementary density functional theory calculations and molecular dynamics simulations provided insight into the nature of the interaction of charged $C_{60}$ with $CO_2$ as well as trends in the packing of $C_{60}^+$ and $C_{60}^-$. The quadrupole moment of $CO_2$ itself was seen to be decisive in determining the charge dependence of the observed adsorption features. Our findings are expected to apply to adsorption of $CO_2$ by charged surfaces in general.**

## Introduction

The role of $CO_2$ as a greenhouse gas has provided a major driving force for experimental and theoretical studies of $CO_2$ adsorption and sequestration in mesoporous materials [1], metal organic frameworks and materials [2,3], zeolites [4,5], and polymer composite materials [6]. Pristine and calcium-doped buckminsterfullerene [7] and various other carbonaceous materials also have been explored previously in this context, including graphites [8,9], graphenes [10], nanotubes [11-13] and carbon nanoscrolls [14]. Density functional theory (DFT) has demonstrated that adsorption and also sequestration of $CO_2$ on boron nitride nanosurfaces can be strongly enhanced by surface charges [15,16]. Trinh et al. studied charge dependent $CO_2$ adsorption in carbon mesopores by molecular dynamics simulations and also found enhanced selectivity for $CO_2$ adsorption compared to $H_2$ adsorption near artificially introduced surface charges [17]. In a DFT study of nitrogen doped carbon nanotubes, Jiao et al. obtained a strong charge-dependence for $CO_2$ capture [18].

Here we have chosen buckminsterfullerene, $C_{60}$, as the carbonaceous adsorbate surface. We report the first experimental evidence for the influence of both positive and negative charge on the adsorption capacity of $CO_2$. We have extensive experience with experiments involving the adsorption of a variety of non-polar and some polar molecules on cationic $C_{60}$, $C_{70}$ and their aggregates at ultra-low temperatures (0.37 K) [19-23]. We have reported previously the mass spectrometric observation of remarkable anomalies in $CO_2$ coverage for the $C_{60}$ dimer and trimer cations $(C_{60})_2^+$ and $(C_{60})_3^+$ [24]. Here we track the adsorption of $CO_2$ on single $C_{60}$ cations and anions in the absence of steric constraints. Interesting and surprising anomalies are observed in this case as well, but we will show that these can be attributed to the unique electron distribution within the $CO_2$ molecule and its interaction with the charge on the $C_{60}$, and so provide new fundamental insights into the adsorption of carbon dioxide on charged carbonaceous surfaces.

The $CO_2$ molecule is roughly cylindrical and its quadrupole moment ($Q_a = -4.278$ D Å[25]) corresponds to slightly negative terminal O atoms that can preferentially respond to a positive charge and to a slightly positive C atom that can preferentially respond to a negative charge on isolated $C_{60}$ ions. The $CO_2$ molecules will be vibrationally and rotationally cold at 0.37 K, the temperature of the helium droplet environment in which the adsorption takes place in our experiments.

## Results and discussion

We exposed cold superfluid helium nanodroplets, doped with $C_{60}$ and $CO_2$, to electron ionization and then sampled the ions that are born within the helium nanodroplets and emerge upon droplet evaporization or ejection from the helium nanodroplets. The raw high-resolution mass-spectra of the cations and anions comprising $C_{60}(CO_2)_n^+$ and $C_{60}(CO_2)_n^-$, are shown in Figure 1. Several remarkable and reproducible features are evident from these spectra. The $C_{60}$ cations and anions have the capacity to adsorb a very large number of $CO_2$ molecules within the superfluid helium droplets. Anomalously high intensity peaks emerge at specific values of n, viz. "magic numbers", that imply special stabilities for these cluster ions. Furthermore, and even more remarkable, *the adsorption capacities of $C_{60}$ cations and anions are different being higher for cations than anions,* as can be deduced from the tails of the shown mass spectra for high mass to charge ratio.

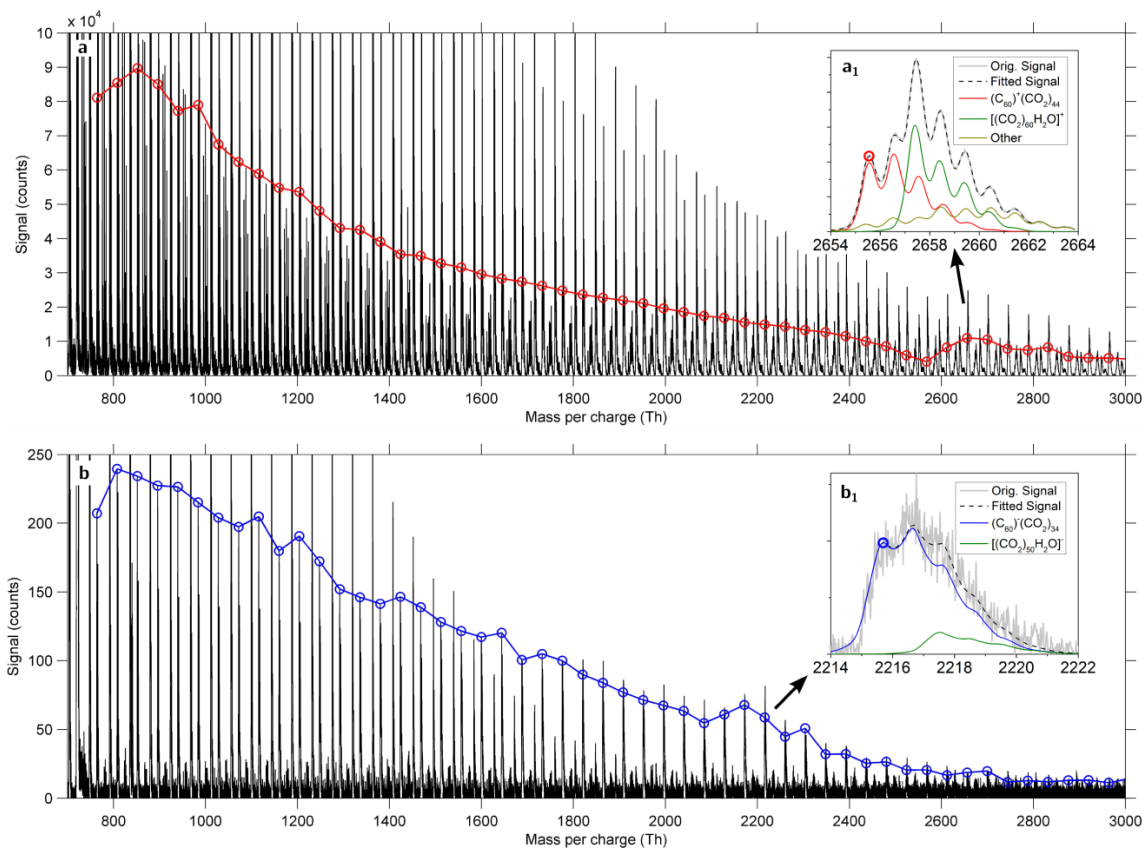

Figure 1: Raw mass spectrum for a) $(C_{60})^+(CO_2)_n$ and b) $(C_{60})^-(CO_2)_n$. The red and blue lines show the highest ion signal for the first isotopologue of the corresponding $C_{60}^\pm(CO_2)_n$ cluster. These data points are not corrected for contributions of other species that may overlap with the $(C_{60})^+(CO_2)_n$ signal. The insets $a_1$ and $b_1$ show fitted ion distribution curves of different species in the mass range of $(C_{60})^+(CO_2)_{44}$ ($a_1$) and $(C_{60})^-(CO_2)_{34}$ as interpreted by our evaluation software IsotopeFit [26].

We approached the interpretation of these anomalous spectral features by means of computations, both with DFT and molecular dynamics (MD) simulations. To gain insight into the interactions of a single $CO_2$ molecule with $C_{60}$, $C_{60}^+$ and $C_{60}^-$, DFT computations were performed. Some of the preferred optimized structures together with their geometrical parameters are shown in Figure 2. Equilibrium dissociation energies ($D_e$, also referred to as adsorption energies) are recorded in Table 1. Equilibrium refers to optimized geometries for both the compound and the dissociated products. We also show the dissociation energies ($D_{MD}$) calculated with the classical force field used in the MD calculations for DFT optimized geometries. As can be expected for shallow potential energy surfaces and the harmonic approximation for vibrational frequency calculations, not all structures in Table 1 are true local minima if no symmetry constraints are present. For example, the vertical orientations over pentagons and hexagons on neutral $C_{60}$ correspond to transition structures of $2^{nd}$ ($N_f = 2$) and $3^{rd}$ ($N_f = 3$) order respectively. With the addition of diffuse basis functions for anion frequency calculations could not be finished, even though the optimizations converged, so that for anions the reported structures might not be true minima. Although the computed adsorption energy of a single $CO_2$ molecule is slightly higher for anions than for cations, in this work we show that the $CO_2$-adsorption capacity in the first adsorption shell is higher for cations than for anions due to steric effects.

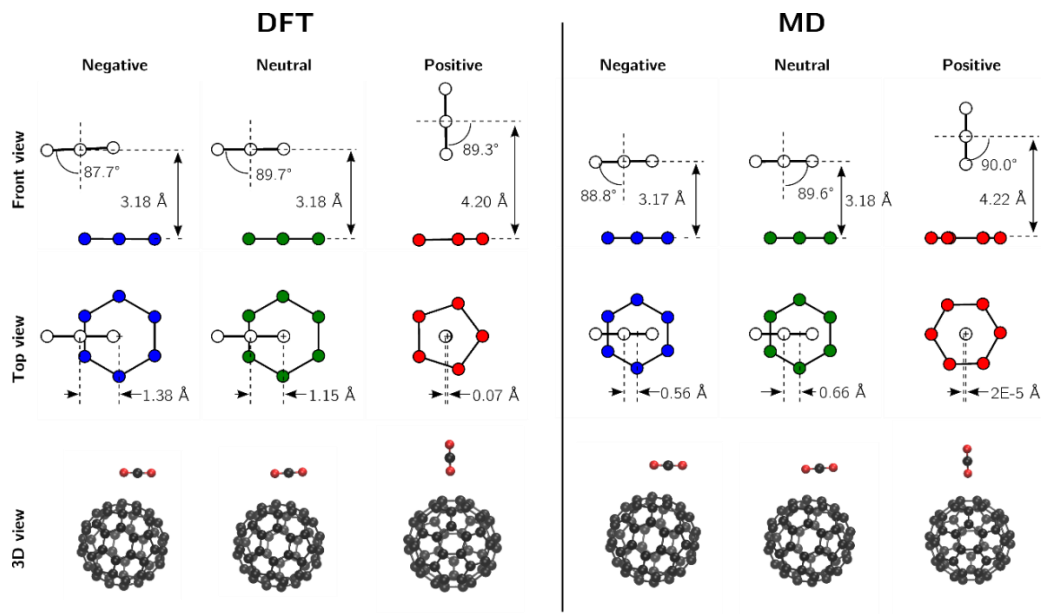

Figure 2. Front and top views of the three preferred orientations of a single $CO_2$ molecule adsorbed on $C_{60}^-$, $C_{60}$ and $C_{60}^+$ as indicated by DFT and MD calculations.

Table 1. Total energies for the adsorption of a single $CO_2$ molecule on $C_{60}$, $C_{60}^+$ and $C_{60}^-$ from DFT calculations. The preferred structures shown in Figure 2 are highlighted.

| | q/e | basis | site | orien | T/eV | ZP/eV | $N_f$ | CP/eV | $D_e$/eV | $D_0$/eV | $D_{CP}$/eV | $D_{MD}$/eV |
|---|---|---|---|---|---|---|---|---|---|---|---|---|
| $C_{60}$ | 0 | 6-31g(d) | | | -62190.423 | 10.435 | | | | | | |
| $C_{60}$ | 1 | 6-31g(d) | | | -62182.757 | 10.354 | | | | | | |
| $C_{60}$ | -1 | 6-31g(d) | | | -62193.698 | | | | | | | |
| $CO_2$ | 0 | 6-31g(d) | | | -5129.890 | 0.322 | | | | | | |
| $CO_2$ | 0 | 6-31g+(d) | | | -5130.079 | 0.321 | | | | | | |
| **$C_{60}CO_2$** | **0** | **6-31g(d)** | **hex** | **flat** | **0 (-67320.4330)** | **10.759** | **0** | **0.050** | **0.120** | **0.118** | **0.070** | **0.100** |
| $C_{60}CO_2$ | 0 | 6-31g(d) | pent | flat | 0.010 | 10.762 | 0 | | 0.109 | | | 0.094 |
| $C_{60}CO_2$ | 0 | 6-31g(d) | pent | vert | 0.066 | | 2 | | 0.054 | | | 0.054 |
| $C_{60}CO_2$ | 0 | 6-31g(d) | hex | vert | 0.074 | | 3 | | 0.046 | | | 0.060 |
| **$C_{60}CO_2$** | **1** | **6-31g(d)** | **pent** | **vert** | **0 (-67312.7744)** | **10.700** | **0** | **0.029** | **0.127** | **0.103** | **0.098** | **0.083** |
| $C_{60}CO_2$ | 1 | 6-31g(d) | hex | vert | 0.002 | 10.692 | 0 | 0.029 | 0.125 | 0.109 | 0.096 | 0.090 |
| $C_{60}CO_2$ | 1 | 6-31g(d) | hex | flat | 0.015 | | 1 | | 0.112 | | | 0.076 |
| $C_{60}CO_2$ | 1 | 6-31g(d) | pent | flat | 0.015 | 10.695 | 0 | | 0.112 | | | 0.072 |
| **$C_{60}CO_2$** | **-1** | **6-31g+(d)** | **hex** | **flat** | **0 (-67323.9386)** | | | **0.028** | **0.162** | | **0.134** | **0.119** |
| $C_{60}CO_2$ | -1 | 6-31g+(d) | pent | flat | 0.025 | | | | 0.136 | | | 0.109 |
| $C_{60}CO_2$ | -1 | 6-31g+(d) | pent | vert | 0.104 | | | | 0.057 | | | 0.023 |
| $C_{60}CO_2$ | -1 | 6-31g+(d) | hex | vert | 0.113 | | | | 0.049 | | | 0.031 |

| | |
|---|---|
| q | Charge |
| T | Total energy or relative energy |
| ZP | Zero point correction |
| $N_f$ | Number of imaginary frequencies (0: local minimum, >0: transition structures) |
| CP | Counterpoise correction |
| De (eV) | dissociation energy from uncorrected total energies |
| $D_0$(eV) | dissociation energy from zero point corrected energies |
| $D_{CP}$ (eV) | dissociation energy from counterpoise corrected energies |
| $D_{MD}$ (eV) | dissociation energy from force field of DFT optimized geometries |
| site | adsorption site |
| orien | orientation of $CO_2$ in the optimized structure |

The results summarized in Table 1 indicate that hexagons are preferred over pentagons as adsorption sites for $C_{60}$ and $C_{60}^-$. The situation is more complex for $C_{60}^+$ with an only slight preference for hexagonal sites if zero-point correction is taken into account ($D_0$). Horizontal adsorption is preferred on $C_{60}$ and $C_{60}^-$ by 0.06 eV or more, while vertical adsorption is preferred on $C_{60}^+$. These preferences can be understood in terms of the relatively positive carbon center (+ 0.76 e in terms of Mulliken charges) and negative oxygen ends of the $CO_2$ molecule (- 0.38 e). The energy difference between the vertical and horizontal configuration (without any corrections) is only 14.5 meV for the cation (vertical preferred), 104 meV for the anion (horizontal preferred), and for the neutral 66 meV (horizontal preferred). The adsorbed $CO_2$ is slightly bent (177.8 °) in the case of the anionic complex. Mulliken charges imply that $CO_2$ becomes slightly polar upon adsorption on $C_{60}^{+,0,-}$ and a small amount of charge is transferred to the fullerene (< 0.2 e). The charge transfer due to $CO_2$ adsorption on $C_{60}$ is visualized in Figure 3 for (a) the cation (b) the neutral (c) the anion. A completely different scheme for charge accumulation (red) and depletion (blue) is observed for the cation and the anion and the induced polarity of $CO_2$ can be clearly seen in panel (a). Note that the isovalues of the three surfaces per panel are small (± 0.0001, ± 0.0002, ± 0.0004 a.u.) because the overall charge transfer is also rather small. In bare $C_{60}^-$ the additional charge is distributed over the whole fullerene (not shown) and resembles the LUMO molecular orbital of $C_{60}$.

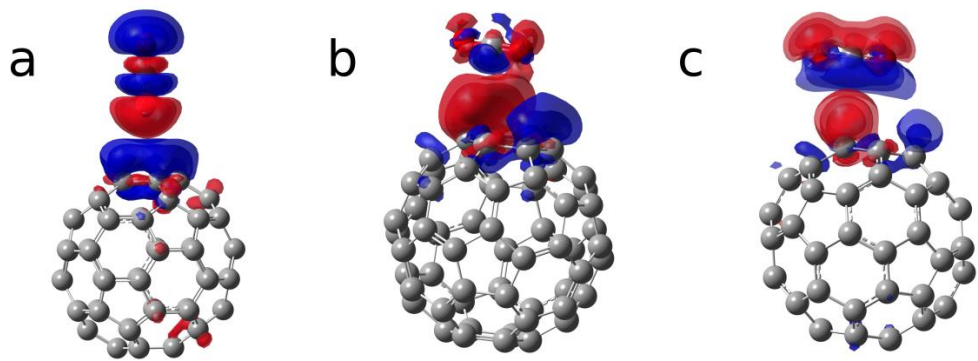

Figure 3: Charge accumulation (red colors) and depletion (blue colors) due to $CO_2$ adsorbed on (a) $C_{60}^+$, (b) $C_{60}$, (c) $C_{60}^-$. Surfaces of the charge density difference $\rho(C_{60}CO_2^{+,0,-}) - \rho(C_{60}^{+,0,-}) - \rho(CO_2)$ for isovalues of $\pm\,0.0001$, $\pm\,0.0002$, and $\pm\,0.0004$ (atomic units) are shown.

The computed binding energies and binding modes can be compared with previously published results for $CO_2$ adsorption on carbonaceous materials. In the classical model of Arab et al. $CO_2$ adsorbs horizontally on external sites of a single walled nanotube (SWNT) bundle with a binding energy of 0.113 eV [12]. This binding energy is very close to our estimate of 0.120 eV on neutral $C_{60}$. The energy induced by carbon polarizability can be neglected [12]. The $CO_2$ dimer interaction energy on SWNT bundles lies between 55 and 64 meV depending on the adsorption site. A parallel, slightly shifted configuration was found except for interstitial sites. In MP2/AVTZ calculations adsorption energies of 0.117 eV on graphite and 0.113 eV on a (9,0) nanotube exceed the experimental heat of adsorption of 0.024 eV in SWNTs [11]. As in our results for neutral $C_{60}$, $CO_2$ favors the horizontal adsorption above the center of a $C_6$ ring. On neutral $C_{60}$ a $CO_2$ adsorption energy of only 0.037 eV was reported employing GGA/PBE density functional theory without dispersion correction. However, the optimized geometry resembles our results. Gao et al. also showed that adsorption can be enhanced by calcium doping [7].

Another GGA/PBE calculation yielded a binding energy of 0.400 eV for $CO_2$ adsorbed on graphene at a distance of 3.0 Å. The vertical configuration adsorbs only with 0.294 eV [27]. In plane-wave DFT $CO_2$ adsorbs on graphene nanoribbons with 0.31 eV binding energy [10]. On SWNTs LDA/PW calculations yielded adsorption energies between 0.089 and 0.109 eV with $CO_2$ positioned between 3.23 and 3.54 Å above hexagons [28]. Reactions pathways of the dissociative adsorption of $CO_2$ on graphite were investigated by Xu et al. [8] A smallest barrier of 5.17 eV was found for dissociative adsorption of a single oxygen atom. The reaction is endothermic with 4.2 eV. A combined experimental and theoretical study of $CO_2$ chemisorption on carbonaceous surfaces revealed a first region of high and decreasing adsorption energy and a second region of high coverage, where the adsorption energy stabilizes in the range of 0.217-0.39 eV [29].

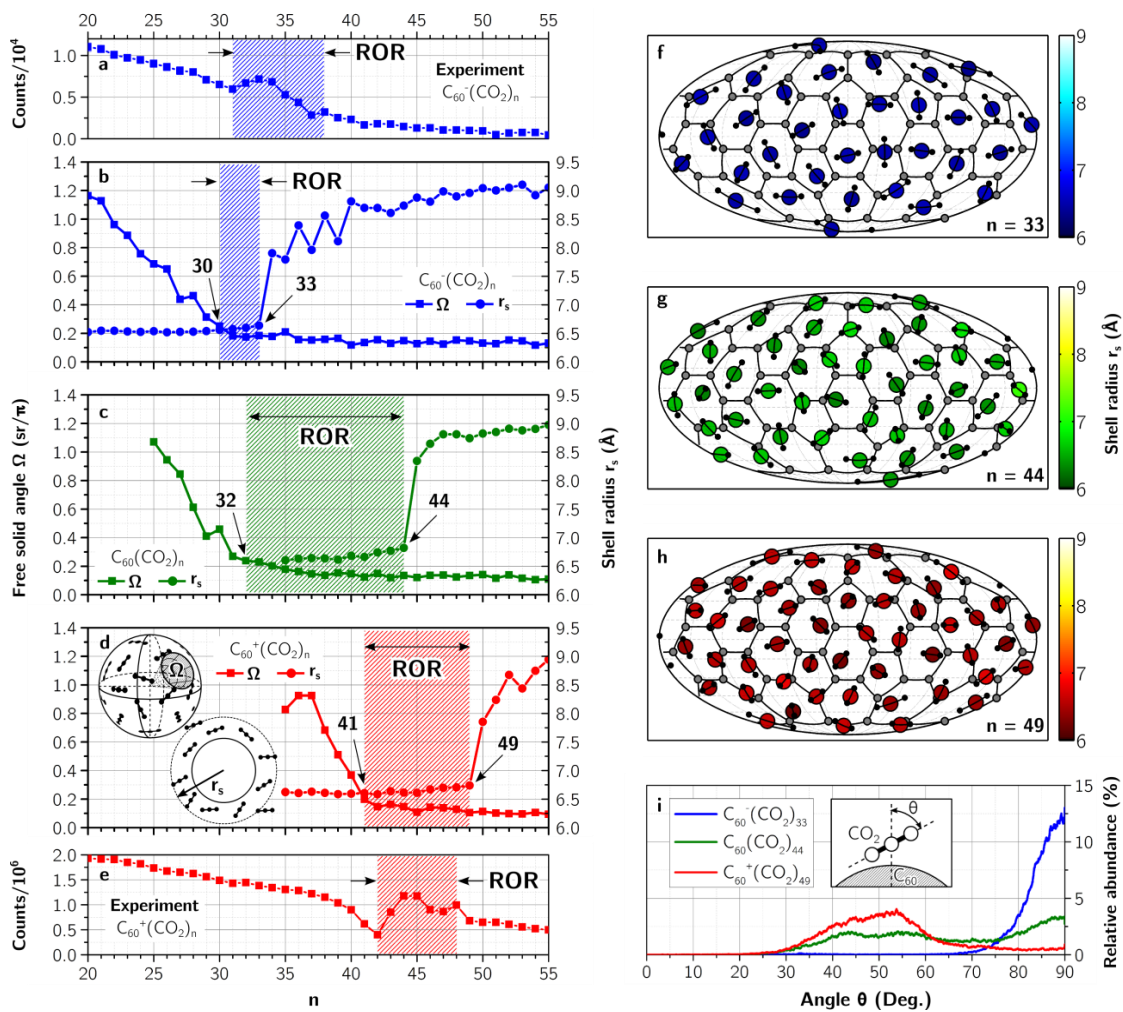

Figure 4: Ion yield measurements in the region of the first shell closure (a, e) and results from MD simulations (b - d, f – i). The positions of the $CO_2$s on the $C_{60}$ and their angles in the case of full coverage are shown in f-i.

The results of our MD simulations provide atomistic insight into the $CO_2$ layers on neutral cationic and anionic $C_{60}$ with respect to the number of molecules in the first shell and their orientation. To assess the $CO_2$ capacity in the first shell, we computed the "shell radius" as the smallest sphere within which a given number of adsorbed $CO_2$ molecules resides (Figure 4 b-d, f-h). With this approach, we find 33, 44 and 49 $CO_2$ molecules to reside within the first shell for negative, neutral and positive $C_{60}$, respectively. Relative to neutral $C_{60}$, the "storage capacity" in the first shell therefore decreases by 25% for the $C_{60}$ anions and increases by about 10% for its cations. The shell closures for the cation at n = 48 (49 in the MD simulation) and for the anion at n = 38 (33 in the MD simulation) are visible in the experimental results shown in Figure 4e and 4a respectively. As expected from our experience, the MD simulations only give qualitatively reasonable results for the anionic case.

Surface diagrams for fully covered neutral and ionic $C_{60}$, together with the angular distribution functions are shown in Figure 4 f-i. Striking differences in the orientational distributions can be seen which can be explained by the quadrupole moment of the $CO_2$ molecule. Electrostatic interaction between the positive carbon center and a negatively charged $C_{60}$ favors a flat orientation for the $CO_2$ molecules. On the other hand, the electrostatic interaction between the negative terminal O atoms in $CO_2$ (or the combined negative charge of the split charges for oxygen in the model of Murthy[30] in the MD simulations) and a positively charged $C_{60}$ favors a tilted orientation, with $CO_2$ standing up by 40-60 degrees. In neutral $C_{60}$ the alignment of $CO_2$ shows an intermediate behavior. These angular distributions do not completely resemble the DFT/MD results for single $CO_2$ adsorption (perfectly vertical/flat) because of the intermingling $CO_2$-$CO_2$ intermolecular interactions.

An interesting feature was observed in the simulations when looking at the $CO_2$-free solid angle $\Omega$. Omega is defined as the largest possible opening angle in steradians of a cone that has its top in the center of the fullerene and does not contain any centers of the adsorbed $CO_2$ atoms. For different coverage numbers for the three charge states of $C_{60}$ we saw a "bald spot" ($\Omega \gg 0$) that dissipates with increasing coverage numbers in each case (Figure 4 b-d). Figure 5 shows the formation and evolution of this bald spot on charged and neutral $C_{60}$. Clearly, in each case, the bald spots fill out with increasing $CO_2$ adsorption until the surface is fully covered. Then several more molecules fit into the first shell before a second layer begins to form (i.e. full coverage is possible over a certain range of n before a second layer starts to form). In between there is a "region of rearrangement" (ROR: $\Omega<\pi/4$, $r_s<7Å$) in which the first shell molecules change their positions and angles. This region is highly charge-dependent and is revealed in the experiment as a broad anomaly in the cluster size distribution rather than a sharp peak (Figure 4a and 4e).

In the simulations, the free solid angle $\Omega$ almost vanishes at a coverage of n = 30 for the anion and n = 41 for the cation. Thus the first minima in the experimental data at n = 31 for the anion and n = 42 for the cation are interpreted as the closure of the bald spot. The simulations also show that $C_{60}$ is fully covered within a certain range of n (30 to 33 for the anion, 41 to 49 for the cation) due to $CO_2$ rearrangement without starting a second adsorption shell. Thus the last anomalies in the ion yield (the right end of the region of rearrangement ROR) at n = 38 (anion) and n=48 (cation) are interpreted as complete closure of the first adsorption shell. Note that the simulations for the anion predict a much narrower ROR than observed in the experiment, which might be an artifact of the force field or of the fixed charge distribution. However, DFT calculations of MD optimized structures revealed a surprisingly homogeneous distribution of the additional electron of $C_{60}^-$ that resembles the lowest unoccupied orbital of $C_{60}$.

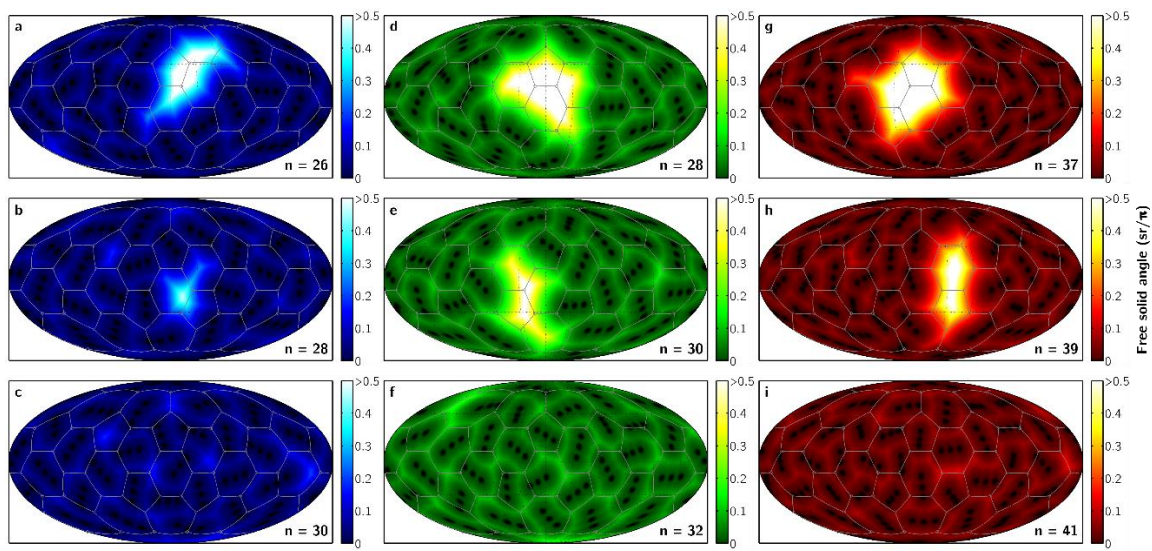

Figure 5. Computed $CO_2$-free solid angle on negatively (a-c), neutral (d-f) and positively (g-i) charged $C_{60}$ surface showing the formation of a bald spot and its dissipation for increasing coverage number n of $CO_2$ molecules.

Our attempts to understand our experimental results with the application of DFT and MD simulations have provided a vivid picture of the step-by-step coverage of $C_{60}$, $C_{60}^+$ and $C_{60}^-$. The electronic properties of $CO_2$ have decisive consequences for the nature of the adsorption as the captured $CO_2$ molecules respond to the absence or presence of charge (positive or negative) on the $C_{60}$ surface. An unambiguous charge order in $CO_2$ adsorption *capacity* was clearly observed experimentally for the first time. Our computations have allowed us to understand this charge order in terms of the orientation of the $CO_2$ molecules on the charged carbonaceous surface that results from the electrostatic interaction between the charge on the surface and the quadrupole of $CO_2$. This orientation effect in the presence of charge is expected to apply generally to adsorption on other charged carbonaceous surfaces, and to surfaces in general, and so to control (enhance or diminish) the capture of $CO_2$ by adsorption.

**Computational details**

The energetics and geometries of neutral, cationic, and anionic $C_{60}CO_2$ structures were calculated by means of density functional theory (DFT). All structures were fully optimized without constraints in order to account for the slight deformation of fullerene and of $CO_2$ arising from each in the presence of the other. Configurations for the adsorption of a single $CO_2$ molecule were optimized, starting from five different initial geometries with OCO flat (two centered over a hexagon (hex), three over a pentagon (pent)) and two initial geometries (one centered over a hexagon, one over a pentagon) for $CO_2$ vertical (vert) adsorption. The calculations were performed with the GAUSSIAN 09 program [31] with the ωB97X-D dispersion corrected hybrid density functional [32, 33] and the 6-31g(d) basis set [34, 35] with additional diffuse functions (+) for anions [36]. We found this density functional to be among the best long-range corrected ones [37] in previous studies on the adsorption of $H_2$, $CH_4$ and $C_2H_4$ on fullerenes [19-22].

MD-simulations were performed to provide structural insight into the observed states of physisorption. Non-polarizable $CO_2$ and $C_{60}$ were kept rigid using the quaternion propagation formalism. We used parameters from Martínez-Alonso et al. [38] for $C_{60}$-$CO_2$ forces and added a distributed charge of ±1 e on the fullerene ions. The charge primarily resides on the fullerene because of its low ionization energy (7.6 eV) [19] and its high electron affinity (2.6 eV) [39] compared to $CO_2$ (13.78 and - 0.6 eV respectively) [25]. Mulliken partial charges from the DFT calculations sum up to 0.95 e on $C_{60}$ for cationic $C_{60}CO_2$ and -1.20 e on $C_{60}$ for anionic $C_{60}CO_2$. The charge distributions of the missing/additional electron resemble the HOMO/LUMO molecular orbitals of $C_{60}$ also for higher $CO_2$ occupations. The small, but non-negligible, influence due to the quadrupole moment of the $CO_2$ adsorbates on the charge distribution cannot be reflected in our MD-simulations. However, the similarities in energetic order (Table 1) and optimized geometries between DFT and MD (Figure 2) results indicate that the model adopted for anions and cations is capable of reproducing the major adsorption features of charge dependent adsorption capacity.

For the $CO_2$-$CO_2$ interaction, we used a 5-charge model (a split charge for each oxygen 0.1216 e/-0.6418 e and one site centered at carbon with 1.0404 e) developed by Murthy et al.[30] that has been used previously to simulate $CO_2$ clusters in vacuum [40, 41]. After initializing a randomly distributed dilute cloud of $CO_2$ molecules around $C_{60}$, the computational procedure consisted of two steps: First, the system was heated to temperatures of 120 K and cooled down again within a total simulation time of 50 ps. Second, the energy was locally optimized with a trust-region optimization algorithm. The annealing towards much higher temperatures than in the experiment guarantees a better sampling of the potential energy surface in finding distinct initial geometries for the local optimizer.

In order to find geometries with low total configuration energy, we inspected a large number of such simulations for each cluster (2500 to 4000, depending on the cluster size). However, due to the size of the system and the huge configurational space, global minima were only reached with very few $CO_2$ adsorbate molecules, so that dissociation energies for larger clusters could not be obtained reliably and were not reported in this work.

Angular distributions were obtained from the $CO_2$ coordinates at all integration steps during the cooling process of longer simulations (200 ps in total with a time step of 500 fs) with lower annealing temperatures (50K to 80K in steps of 10K). For a representation of the shell radius we used the distance between the $C_{60}$ center of mass and the closest atom of the outermost $CO_2$ molecule. We searched for the smallest shell radius in all simulations and integration steps for a given cluster size.

Structures were simulated and optimized with our own code that is specialized for simulations and structure optimizations of clusters in vacuum with flexible force fields and fast automation and data analysis.

**Experimental details**

The experimental setup has been described in detail elsewhere [42, 43]. In brief, helium (purity 99.9999%) is cooled to 9.5K by a closed cycle two-stage cryocooler (SRDK-415D-F50H, Sumitomo Heavy Indus-tries Ltd.). Helium nanodroplets are formed by helium expansion at a stagnation pressure of 2.0 MPa through a 5 μm nozzle into vacuum. Under these conditions the estimated average number of helium atoms per droplet is in the order of $5 \times 10^5$. The droplets are superfluid with a temperature of 0.37K [44]. After formation, the helium droplet beam passes a 0.8 mm conical skimmer to avoid shock waves and enters a differentially pumped pickup chamber. The pickup chamber is again divided into two differentially pumped regions. A small amount of $C_{60}$ (SES research, purity 99.95%) is vaporized into the first region from a heated crucible. In the second region $CO_2$ (Messer; purity 99.9995%) is introduced from an external reservoir and fed into the chamber with a flow controller. Stable and efficient pickup conditions are achieved at a constant source temperature of 330 °C for $C_{60}$ and 2 mPa for $CO_2$. After the pickup process the He beam enters the ionization chamber and is crossed with an electron beam of 8.5 eV and 120 eV for the optimum production of anions and cations, respectively. The resulting anions and cations are guided by a weak electrostatic field toward the entrance of a time-of-flight mass spectrometer. The commercial orthogonal reflectron time-of-flight mass spectrometer (Tofwerk) separates the masses and achieves a mass resolution of R ~5000FWHM (in V-mode). Ultimately the ions are detected by a multi-channel plate operated in a single counting mode.

The evaluation of the raw mass spectra was performed with our own software IsotopeFit, which is able to extract the abundances of all peaks in the spectrum by fitting the ion distribution curves of all species that contribute to the signal [26]. The insets in Figure 1 show the raw spectra for $C_{60}^{-}(CO_2)_{34}$ and $C_{60}^{+}(CO_2)_{44}$ together with the fitted ion distribution curves in the corresponding mass range as an example. Please note the slightly different curves of the raw spectra in Figure 1 and the extracted total counts per ion in Figure 4a and 4e which stems from overlapping ion signals as well as area distortions due to isotopic effects which are eliminated by summation over the fitted signals for contributing isotopes.

## Acknowledgements

This work was given financial support by the Austrian Science Fund (FWF), Wien (P26635, and I978). This work was supported by the Austrian Ministry of Science BMWF as part of the UniInfrastrukturprogramm of the Focal Point Scientific Computing at the University of Innsbruck.

1. T. A. Steriotis, K. L. Stefanopoulos, F. K. Katsaros, R. Gläser, A. C. Hannon and J. D. F. Ramsay, *Phys. Rev. B*, 2008, **78**, 115424.
2. O. Shekhah, Y. Belmabkhout, Z. Chen, V. Guillerm, A. Cairns, K. Adil and M. Eddaoudi, *Nat. Comm.*, 2014, **5**, 4228.
3. B. K. Chang, P. D. Bristowe and A. K. Cheetham, *Phys Chem Chem Phys*, 2013, **15**, 176-182.
4. J. Kim, L.-C. Lin, J. A. Swisher, M. Haranczyk and B. Smit, *J. Am. Chem. Soc.*, 2012, **134**, 18940-18943.
5. I. Matito-Martos, A. Martin-Calvo, J. J. Gutierrez-Sevillano, M. Haranczyk, M. Doblare, J. B. Parra, C. O. Ania and S. Calero, *Phys Chem Chem Phys*, 2014, **16**, 19884-19893.
6. T. Rodenas, I. Luz, G. Prieto, B. Seoane, H. Miro, A. Corma, F. Kapteijn, F. X. Llabrés i Xamena and J. Gascon, *Nat. Mater.*, 2015, **14**, 48-55.
7. B. Gao, J.-x. Zhao, Q.-h. Cai, X.-g. Wang and X.-z. Wang, *J. Phys. Chem. A*, 2011, **115**, 9969-9976.
8. S. C. Xu, S. Irle, D. G. Musaev and M. C. Lin, *J. Phys. Chem. B*, 2006, **110**, 21135-21144.
9. T. T. Trinh, D. Bedeaux, J. M. Simon and S. Kjelstrup, *Phys Chem Chem Phys*, 2015, **17**, 1226-1233.

10. B. Huang, Z. Li, Z. Liu, G. Zhou, S. Hao, J. Wu, B.-L. Gu and W. Duan, *J. Phys. Chem. C*, 2008, **112**, 13442-13446.
11. M. Cinke, J. Li, C. W. Bauschlicher, A. Ricca and M. Meyyappan, *Chem. Phys. Lett.*, 2003, **376**, 761-766.
12. M. Arab, F. Picaud, M. Devel, C. Ramseyer and C. Girardet, *Phys. Rev. B*, 2004, **69**, 165401.
13. M. Bienfait, P. Zeppenfeld, N. Dupont-Pavlovsky, M. Muris, M. R. Johnson, T. Wilson, M. DePies and O. E. Vilches, *Phys. Rev. B*, 2004, **70**, 035410.
14. D. Mantzalis, N. Asproulis and D. Drikakis, *Phys. Rev. E*, 2011, **84**, 066304.
15. Q. Sun, Z. Li, D. J. Searles, Y. Chen, G. Lu and A. Du, *J. Am. Chem. Soc.*, 2013, **135**, 8246-8253.
16. H. Guo, W. Zhang, N. Lu, Z. Zhuo, X. C. Zeng, X. Wu and J. Yang, *J. Phys. Chem. C*, 2015, **119**, 6912-6917.
17. T. T. Trinh, T. J. H. Vlugt, M.-B. Hägg, D. Bedeaux and S. Kjelstrup, *Energy Procedia*, 2015, **64**, 150-159.
18. Y. Jiao, Y. Zheng, S. C. Smith, A. Du and Z. Zhu, *ChemSusChem*, 2014, **7**, 435-441.
19. A. Kaiser, C. Leidlmair, P. Bartl, S. Zöttl, S. Denifl, A. Mauracher, M. Probst, P. Scheier and O. Echt, *J. Chem. Phys.*, 2013, **138**, 074311-074313.
20. S. Zöttl, A. Kaiser, M. Daxner, M. Goulart, A. Mauracher, M. Probst, F. Hagelberg, S. Denifl, P. Scheier and O. Echt, *Carbon*, 2014, **69**, 206-220.
21. S. Zöttl, A. Kaiser, P. Bartl, C. Leidlmair, A. Mauracher, M. Probst, S. Denifl, O. Echt and P. Scheier, *J. Phys. Chem. Lett.*, 2012, **3**, 2598-2603.
22. A. Kaiser, S. Zöttl, P. Bartl, C. Leidlmair, A. Mauracher, M. Probst, S. Denifl, O. Echt and P. Scheier, *ChemSusChem*, 2013, **6**, 1235 – 1244.
23. C. Leidlmair, Y. Wang, P. Bartl, H. Schöbel, S. Denifl, M. Probst, M. Alcamí, F. Martín, H. Zettergren, K. Hansen, O. Echt and P. Scheier, *Phys. Rev. Lett.*, 2012, **108**, 076101.
24. A. Mauracher, A. Kaiser, M. Probst, S. Zöttl, M. Daxner, J. Postler, M. M. Goulart, F. Zappa, D. K. Bohme and P. Scheier, *Int. J. Mass. Spectrom.*, 2013, **354–355**, 271-274.
25. Computation Chemistry and Benchmark DataBase of the National Institute of Standards and Technology, http://cccbdb.nist.gov/, Accessed 2015/01/15.
26. S. Ralser, J. Postler, M. Harnisch, A. M. Ellis and P. Scheier, *Int. J. Mass. Spectrom.*, 2015, **379**, 194-199.
27. A. Ghosh, K. S. Subrahmanyam, K. S. Krishna, S. Datta, A. Govindaraj, S. K. Pati and C. N. R. Rao, *J. Phys. Chem. C*, 2008, **112**, 15704-15707.
28. Z. Jijun, B. Alper, H. Jie and L. Jian Ping, *Nanotechnology*, 2002, **13**, 195.
29. A. Montoya, F. Mondragón and T. N. Truong, *Carbon*, 2003, **41**, 29-39.
30. C. S. Murthy, S. F. O'Shea and I. R. McDonald, *Mol. Phys.*, 1983, **50**, 531-541.
31. M. J. Frisch, G. W. Trucks, H. B. Schlegel, G. E. Scuseria, M. A. Robb, J. R. Cheeseman, G. Scalmani, V. Barone, B. Mennucci, G. A. Petersson, H. Nakatsuji, M. Caricato, X. Li, H. P. Hratchian, A. F. Izmaylov, J. Bloino, G. Zheng, J. L. Sonnenberg, M. Hada, M. Ehara, K. Toyota, R. Fukuda, J. Hasegawa, M. I. a. T., Nakajima, Y. Honda, O. Kitao, H. Nakai, T. Vreven, J. Montgomery, J. A. , J. E. Peralta, F. Ogliaro, M. Bearpark, J. J. Heyd, E. Brothers, K. N. Kudin, V. N. Staroverov, R. Kobayashi, J. Normand, K. Raghavachari, A. Rendell, J. C. Burant, S. S. Iyengar, J. Tomasi, M. Cossi, N. Rega, J. M. Millam, M. Klene, J. E. Knox, J. B. Cross, V. Bakken, C. Adamo, J. Jaramillo, R. Gomperts, R. E. Stratmann, O. Yazyev, A. J. Austin, R. Cammi, C. Pomelli, J. W. Ochterski, R. L. Martin, K. Morokuma, V. G. Zakrzewski, G. A. Voth, P. Salvador, J. J. Dannenberg, S. Dapprich, A. D. Daniels, Ö. Farkas, J. B. Foresman, J. V. Ortiz, J. Cioslowski and D. J. Fox, *Gaussian Inc. Wallingford CT 2009*.
32. S. Grimme, *J. Comp. Chem.*, 2006, **27**, 1787-1799.
33. J.-D. Chai and M. Head-Gordon, *Phys. Chem. Chem. Phys.*, 2008, **10**, 6615-6620.

34. R. Ditchfield, W. J. Hehre and J. A. Pople, *J. Chem. Phys.*, 1971, **54**, 724-728.
35. W. J. Hehre, R. Ditchfield and J. A. Pople, *J. Chem. Phys.*, 1972, **56**, 2257-2261.
36. T. Clark, J. Chandrasekhar, G. W. Spitznagel and P. V. R. Schleyer, *J. Comput. Chem.*, 1983, **4**, 294-301.
37. J. Klimes and A. Michaelides, *J. Chem. Phys.*, 2012, **137**, 120901.
38. A. Martínez-Alonso, J. M. D. Tascón and E. J. Bottani, *J. Phys. Chem. B*, 2000, **105**, 135-139.
39. O. V. Boltalina, L. N. Sidorov, A. Ya. Borshchevsky, E. V. Sukhanova and E. V. Skokan, *Rapid Communications in Mass Spectrometry*, 1993, **7**, 1009-1011.
40. J.-B. Maillet, A. Boutin, S. Buttefey, F. Calvo and A. H. Fuchs, *J. Chem. Phys.*, 1998, **109**, 329-337.
41. J.-B. Maillet, A. Boutin and A. H. Fuchs, *J. Chem. Phys.*, 1999, **111**, 2095-2102.
42. L. An der Lan, P. Bartl, C. Leidlmair, H. Schöbel, R. Jochum, S. Denifl, T. D. Märk, A. M. Ellis and P. Scheier, *J. Chem. Phys.*, 2011, **135**, 044309.
43. H. Schöbel, P. Bartl, C. Leidlmair, S. Denifl, O. Echt, T. D. Märk and P. Scheier, *Eur. Phys. J. D*, 2011, **63**, 209-214.
44. J. P. Toennies and A. F. Vilesov, *Angew. Chem. Int. Ed.*, 2004, **43**, 2622-2648.